\documentclass[twoside,12pt]{article}
\usepackage{epsfig} \date{}
\begin{document}
\newcommand{\nuc}[2]{$^{#2}${#1}}
\newcommand{\heag}{\nuc{He}{3}($\alpha$,$\gamma$)\nuc{Be}{7}}

\newcommand{\op}[1]{#1}
\newcommand{\ket}[1]{\big| {#1} \big> }
\newcommand{\braket}[2]{\big< {#1} \big| {#2} \big> }
\title{ \vspace{1cm}Towards Microscopic Ab Initio Calculations\\ of Astrophysical $S$-Factors$^\dagger$}

\author{Thomas Neff, Hans Feldmeier and Karlheinz Langanke\\
        GSI Helmholtzzentrum f{\"u}r Schwerionenforschung GmbH\\ 
         Planckstra{\ss}e 1, 64291 Darmstadt, Germany       }
\maketitle
\begin{abstract}
Low energy capture cross sections are calculated within a microscopic many-body approach
using an effective Hamiltonian derived from the Argonne~V18 potential.
The dynamics is treated within Fermionic Molecular Dynamics (FMD) which uses a Gaussian 
wave-packet basis to represent the many-body states. 
A phase-shift equivalent effective interaction derived within the Unitary Correlation Operator 
Method (UCOM) that treats explicitly short-range central and tensor correlations is employed.
As a first application the $^3$He($\alpha$,$\gamma$)$^7$Be reaction
is presented. Within the FMD approach the microscopic many-body wave functions of the
$3/2^-$ and $1/2^-$ bound states in $^7$Be as well as the many-body scattering states
in the $1/2^+$, $3/2^+$ and $5/2^+$ channels are calculated as eigenstates of the
same microscopic effective Hamiltonian. Finally the $S$-factor is calculated from
$E1$ transition matrix elements between the many-body scattering and bound states.  
For \heag\/ the $S$ factor agrees very well, both in absolute normalization and 
energy dependence, with the recent experimental data from the Weizmann, LUNA, Seattle 
and ERNA experiments.
For the {\nuc{H}{3}($\alpha$,$\gamma$)\nuc{Li}{7}} reaction the 
calculated $S$-factor is about 15\% above
the data.
\end{abstract}

\section{Introduction}

Low energy nuclear reactions play an important role in many astrophysical scenarios. 
If the environmental temperature is so low that the kinetic energy of the nuclei
is small compared to the Coulomb barrier, quantum tunneling 
leads to reaction rates which are exponentially dropping with decreasing energy.
Hence it becomes increasingly difficult to impossible to measure the tiny cross sections 
at such small energies and one has often to rely on extrapolations 
to astrophysically relevant energies.
Typically these reactions are described with potential models
where the reaction partners are treated as point-like nuclei interacting
via nucleus-nucleus potentials fitted to experimental data on bound
and scattering states. In a microscopic \emph{ab initio} picture however, the system is
regarded as a many-body system of interacting nucleons. The many-body wave
functions are fully antisymmetrized and realistic nucleon-nucleon
interactions are used.

In the Fermionic Molecular Dynamics (FMD) approach \cite{fmd08,fmd00,fmd97,fmd95} we aim at a
consistent description of bound states, resonances and scattering
states using realistic low-momentum nucleon-nucleon
interactions obtained in the Unitary Correlation Operator Method (UCOM)
\cite{ucom10,ucom06,ucom03,ucom98,fmd04}. 
Intrinsic many-body basis states are Slater determinants
built with Gaussian wave packets as single-particle states. This basis
contains as special cases harmonic oscillator shell model and Brink-type cluster wave
functions. The symmetries of the system are restored by projection on parity, angular momentum 
and total linear momentum. 
The many-body eigenstates of the realistic Hamiltonian are obtained in multi-configuration
mixing calculations. 
FMD has already been used successfully to describe the structure of nuclei in the 
$p$- and $sd$-shell, like the Hoyle state in \nuc{C}{12} \cite{hoyle07} or halo- and 
cluster-structures in Neon isotopes \cite{geithner08}.

In this contribution we extend FMD to the continuum where
many-body states have to be matched to phase shifted Coulomb solutions
of point like nuclei.
We present first results for the \heag\/ and 
\nuc{H}{3}($\alpha$,$\gamma$)\nuc{Li}{7} capture cross sections 
using the FMD approach.
The  \heag\/ reaction, which is the onset of the \nuc{Be}{7} and \nuc{B}{8}
branches of the pp II and pp III chain of hydrogen burning, has been studied theoretically 
since long either using potential models \cite{kim81,mohr09} or microscopic 
cluster models \cite{langanke86,kajino86} assuming \nuc{He}{3}+\nuc{He}{4} cluster wave functions.
Polarization effects were considered by including the \nuc{Li}{6}+$p$ channel in
\cite{mertelmeier86,csoto00}. However,
consistent \emph{ab initio} calculations starting from realistic interactions have not 
been possible up to now. First attempts using Variational Monte-Carlo \cite{nollett01} and 
the No-Core Shell Model \cite{navratil07b} only calculated the asymptotic normalization 
coefficients from the bound state wave functions.

\section{Fermionic Molecular Dynamics}

In FMD the intrinsic many-body basis states are Slater determinants
\begin{equation}
  \label{eq:fmdslaterdet}
  \ket{Q} = \op{\mathcal{A}} \bigl\{ \ket{q_1} \otimes \ldots \otimes
    \ket{q_A} \bigr\} \: ,
\end{equation}
using Gaussian wave packets as single-particle states
\begin{equation}
  \braket{\vec{x}}{q_k} =
  \exp \biggl\{ -\frac{(\vec{x} -\vec{b}_k)^2}{2 a_k} \biggr\} \otimes
  \ket{\chi^\uparrow_{k},\chi^\downarrow_{k}} \otimes \ket{\xi_k} \: .
\end{equation}
The variational set $Q$ contains the complex parameters $\vec{b}_k$, which encode 
the mean positions and momenta of the wave packets, 
the complex width parameters $a_k$, and the spin directions are controlled by
$\chi^\uparrow_{k}$ and $\chi^\downarrow_{k}$. Proton and neutron are 
distinguished by $\xi_k$.
To restore the symmetries of the Hamiltonian the intrinsic wave function $\ket{Q}$ 
is projected on parity, angular momentum and total linear momentum $\vec{P}=0$.
\begin{equation}\label{eq:Qproj}
  \ket{Q; J^\pi MK}\otimes\ket{\vec{P}=0}_{cm} =
  \op{P}^{J}_{MK} \op{P}^\pi \op{P}^{\vec{P}=0}\ \ket{Q} 
\end{equation}
\begin{figure}[b]
	\centering
  \includegraphics[width=0.4\textwidth]{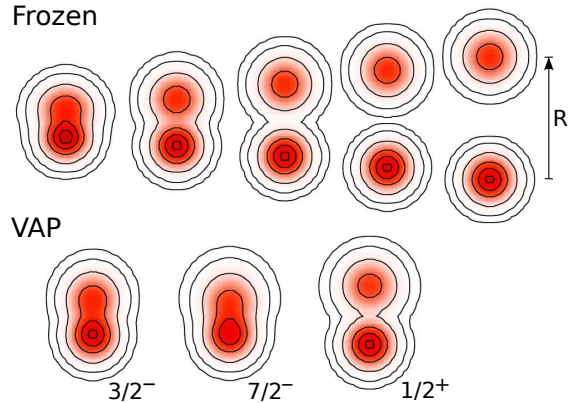}
	\caption{Density distributions of the intrinsic basis states. Top: frozen cluster configurations, 
     bottom: selected polarized configurations obtained by Variation After parity and 
      angular momentum Projection.}
  \label{fig:intrinsic}
\end{figure}

We divide the Hilbert space into an outer region where the system is represented as 
$^3$He and $^4$He nuclei in their ground states interacting only via the Coulomb interaction.
In the inner region the nuclear force acts among all nucleons and
the strong interactions between the approaching nuclei will polarize them.
Therefore one has to enlarge the many-body Hilbert space beyond the microscopic cluster 
model \cite{langanke94} where the Hilbert space is spanned by so called ``frozen'' 
configurations that are just antisymmetrized products of the \nuc{He}{4} and \nuc{He}{3} 
clusters in their ground state placed at various distances $R$, see Fig.~\ref{fig:intrinsic}.

In order account for polarized clusters we generate additional configurations
by a variation after parity and angular momentum projection procedure (VAP). 
The energy of the projected many-body state is minimized with respect to the 
parameters of all single particle states of the intrinsic Slater determinant in the
sub-manifold defined by a fixed mean square radius. 
By dialing the value of the radius constraint from large to small we obtain polarized clusters
down to shell model like configurations. The polarized clusters are essential
for the continuum scattering states and the surface region of the bound states.
The shell model like configurations contribute mainly to bound states.
Some polarized configurations are shown in Fig.~\ref{fig:intrinsic} (VAP). 
The many-body Hilbert space is spanned by the non-orthogonal
projected configurations given in Eq.~(\ref{eq:Qproj}).
The task is to find a set of intrinsic states $\ket{Q^{(a)}}$ that describes the
physics well. Typically, we employ of the order of 50 intrinsic states which are taken
to be the same set for bound and scattering states, of course projected on the
respective $J^\pi$. 

The effective Hamiltonian is diagonalized in the available Hilbert space
with the proper boundary conditions for either bound (discrete values of
the energy $E_n$) or scattering states  ($E=k^2/(2\mu)$).
\begin{equation}
    \op{H}^{\rm{eff}}\ \ket{{J^\pi M, E_n}} = E_n\  \ket{{J^\pi M, E_n}}
\end{equation}
This determines the multi-configuration mixing coefficients $c^{(n)}_{aK}$
of the microscopic many-body eigenstates   
\begin{equation}
      \ket{{J^\pi M, E_n}} =  \sum_{aK} c^{(n)}_{aK}\   
           \ket{Q^{(a)}; J^\pi MK}    \ .
\end{equation}
The effective Hamiltonian $\op{H}^{\rm{eff}}=\op{T}+\op{V}_{\rm{UCOM}}$ is obtained 
within the Unitary Correlation Operator Method (UCOM) by transforming the realistic
Argonne-V18 Hamiltonian in the two-body space \cite{ucom10}.  
This incorporates the short-range correlations into the effective interaction
$\op{V}_{\rm{UCOM}}$ and at the same time leaves the phase shifts of the nucleon-nucleon 
scattering problem and the deuteron energy unchanged.

\section{Bound and Scattering States}

The frozen states can be rewritten using RGM basis states \cite{baye77,descouvemont10}.
\begin{equation}
\ket{\Phi(r);[\,\ell\ \textstyle{\frac{1}{2}}^+]^{J^\pi}_M}= 
        {\cal A}\ \left[ \ket{r,\ell}_{rel}\otimes  
               \ket{^3\rm{He};\textstyle{\frac{1}{2}}^+}\otimes
               \ket{^4\rm{He};0^+}\right]^{\,J^\pi}_M\ ,
\end{equation}
where $\ket{r,\ell}_{rel}$ denotes a state of relative motion with angular momentum $\ell$
in which the center of masses of the internal ground states 
$\ket{^3\rm{He};\textstyle{\frac{1}{2}}^+}$ and 
$\ket{^4\rm{He};0^+}$ are positioned at relative distance $r$. 
Internal and relative motion angular momenta are coupled to total $J$.
The RGM representation can then be used to match the logarithmic derivative of the relative 
wave function of the clusters to the asymptotic Whittaker function for bound states or to 
Coulomb scattering wave functions with phase shifts: 
\begin{equation}
\braket{\Phi(r);[\,\ell\ \textstyle{\frac{1}{2}}^+]^{J^\pi}_M}{\,J^\pi M,E}
        \ \ \stackrel{r\rightarrow\infty}{\longrightarrow}\ \
        \frac{1}{r}\left(\
        F_\ell(k r) + \tan\big(\delta_\ell^{J^\pi}(k)\big)\ G_\ell(k r)\right),
       \ \ \ k=+\sqrt{2\mu E} \ .
\end{equation}
Using the microscopic $R$-matrix approach of the Brussels group \cite{baye77,descouvemont10} 
we solve the Schr{\"o}dinger equation with the proper boundary conditions for 
bound and scattering states.
 
\begin{figure}[h]
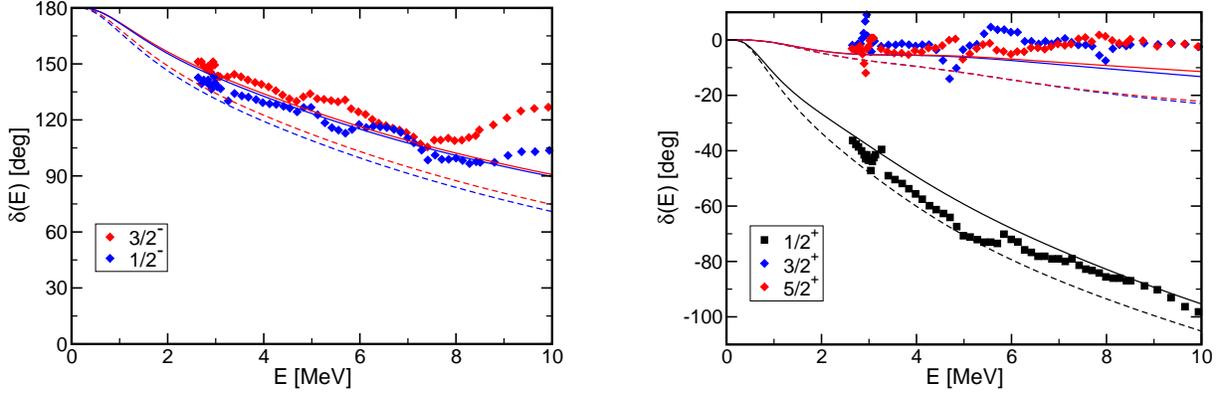

\begin{center}
  \includegraphics[width=0.4\textwidth]{FMD-He4-He3-Phaseshifts-31-.eps}\hfil
	\includegraphics[width=0.4\textwidth]{FMD-He4-He3-Phaseshifts-135+.eps}
\end{center}
	\caption{\nuc{He}{4} + \nuc{He}{3} scattering phase shifts for $P$-waves left and the 
     $S$ and $D$-waves right. Dashed lines show results using only frozen cluster configurations, 
     solid lines show results for the full model space.}
	\label{fig:phaseshifts}
\end{figure}

With frozen configurations alone the $3/2^-$ and $1/2^-$ states in $^7$Be
are bound by only 240~keV and 10~keV,
respectively. In the full Hilbert space, including the polarized configurations, we obtain 
binding energies of 1.49~MeV and 1.31~MeV with respect to the cluster threshold. 
In other words we reproduce the centroid energy but underestimate the splitting of the $3/2^-$ 
and $1/2^-$ states. By artificially increasing the spin-orbit strength we observed that
the total cross section depends essentially only on the centroid energy
while the increased splitting affects only the branching slightly.
The calculated $^7$Be charge radius of 2.67~fm is in good agreement with the experimental value of 
2.647(17)~fm \cite{noertershaeuser09}. 
In Fig.~\ref{fig:phaseshifts} we present the calculated phase shifts together with the 
experimental data \cite{spiger67}. Again we see a sizable effect when we compare the results 
using only the frozen configurations with the results obtained in the full Hilbert space 
including polarization effects.

\section{Spectroscopic amplitudes}

\begin{figure}[b]
  \includegraphics[width=0.245\columnwidth]{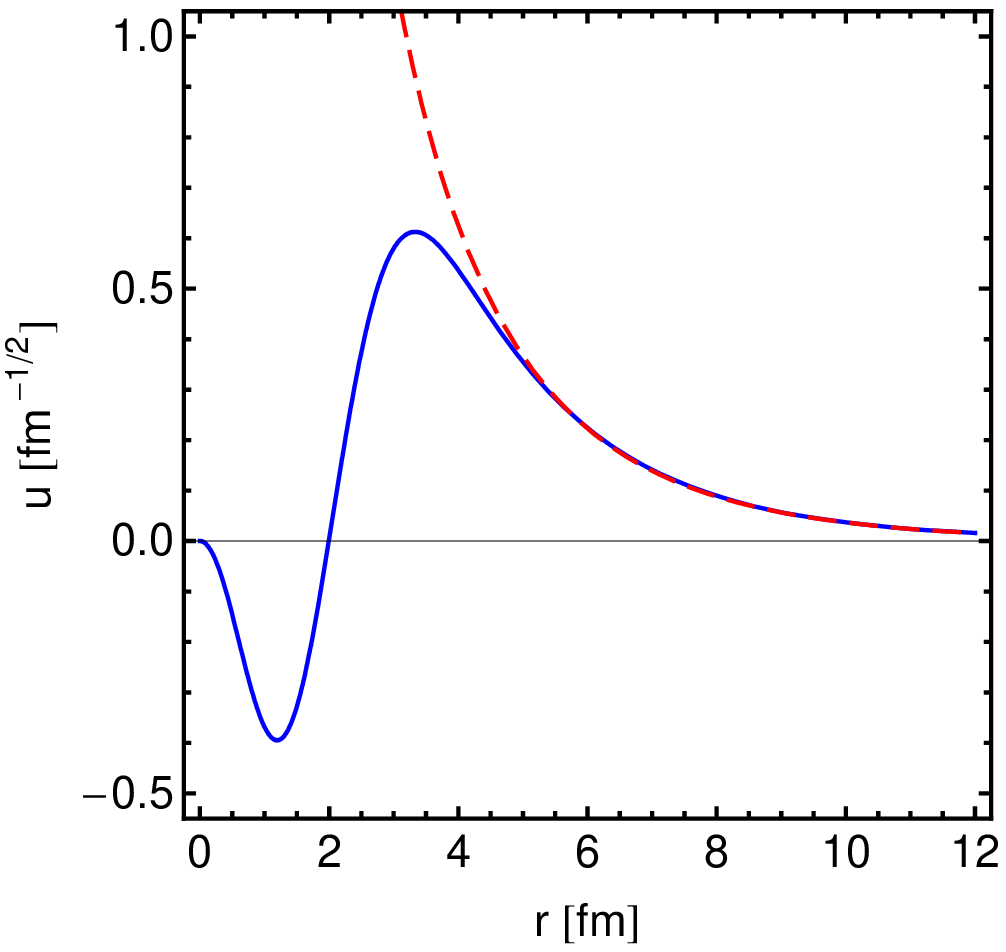}\hfill
  \includegraphics[width=0.253\columnwidth]{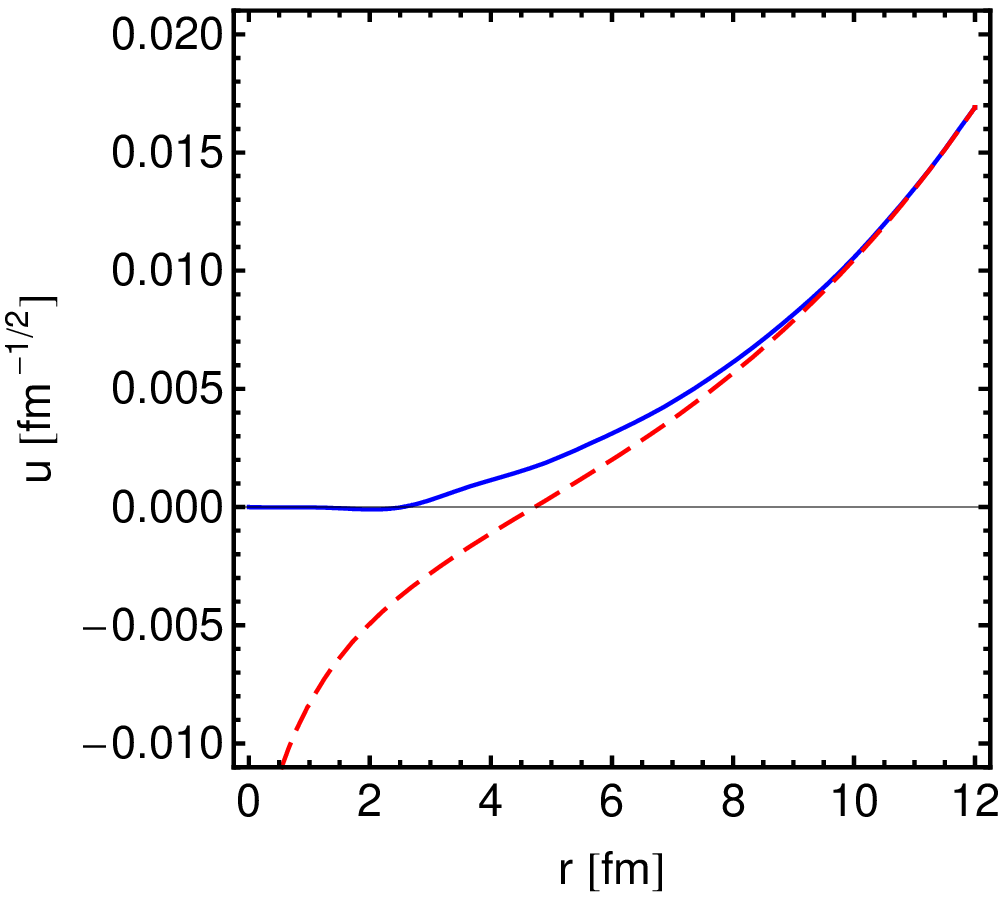}\hfill
  \includegraphics[width=0.245\columnwidth]{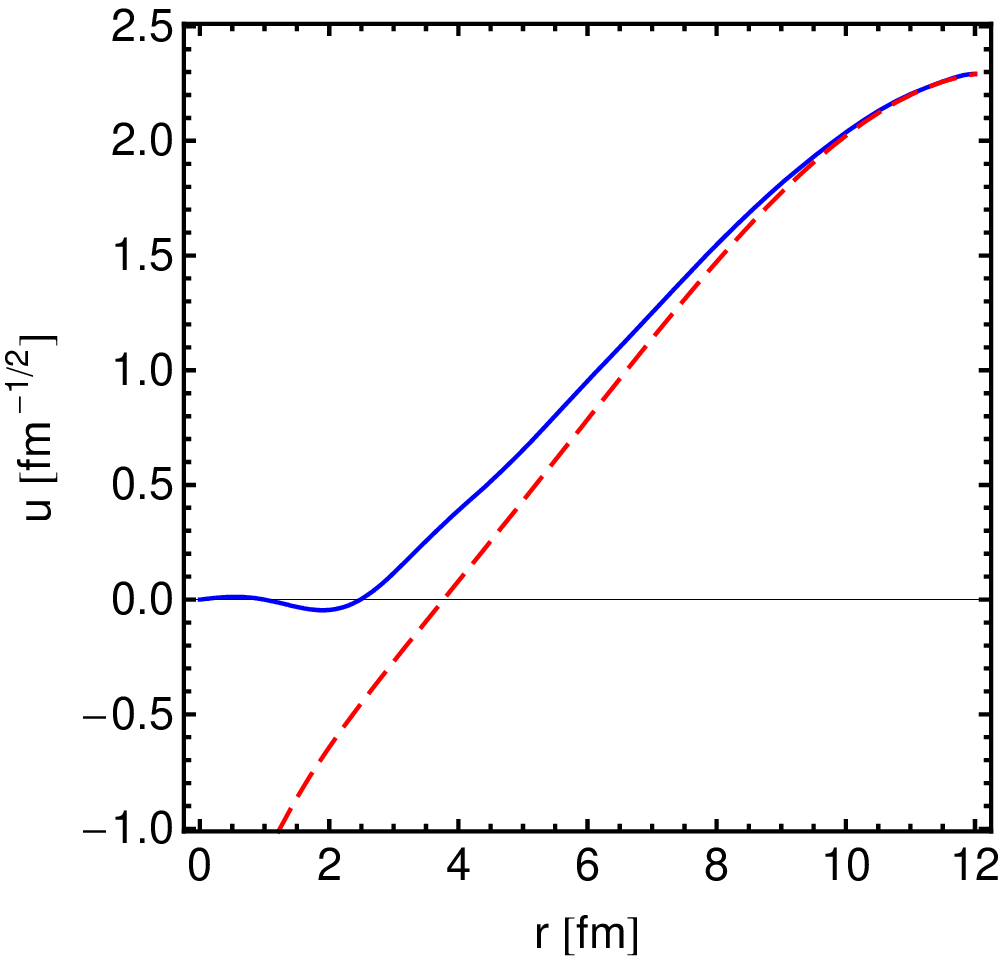}\hfill
  \includegraphics[width=0.240\columnwidth]{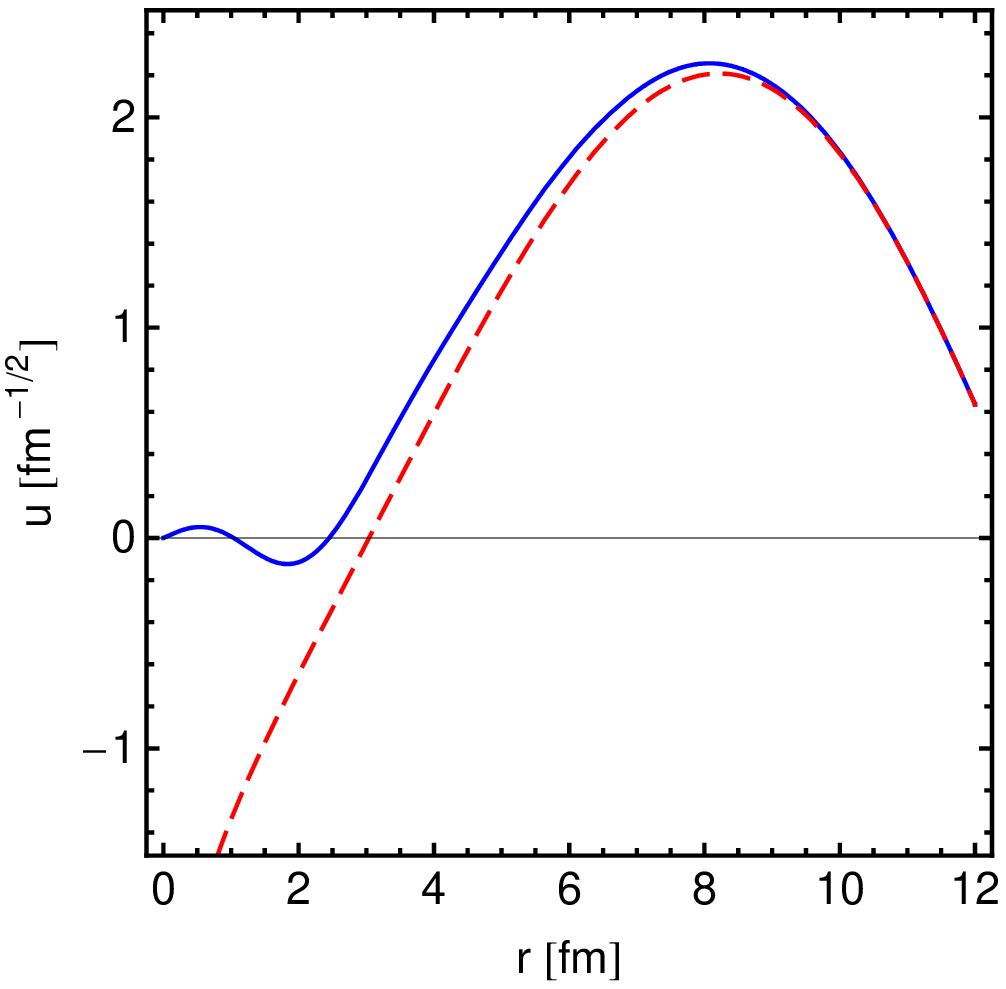}
  \caption{From left to right: overlap functions $r\hat{\psi}^{J^\pi}_E(r)$
   for the relative motion of the clusters for the $3/2^-$ ground state 
    and $1/2^+$ scattering states at energies of 0.1 MeV, 1.0 MeV and 2.0 MeV. 
    Dashed lines show the Whittaker and Coulomb functions matched at the channel radius $a$=12~fm,
    respectively.}
  \label{fig:amplitudes}
\end{figure}
If the many-body system has a pronounced cluster structure, as is the case in
\nuc{Li}{7}, one can define an overlap function $\hat{\psi}^{J^\pi}_E(r)$ for the relative 
motion of the clusters by means of the spectroscopic amplitude 
$\braket{\Phi(r)}{J^\pi\!,E}$ $\equiv$
$\braket{\Phi(r);[\,\ell\ \textstyle{\frac{1}{2}}^+]^{J^\pi}_M}{\,J^\pi M,E}$
for a given $J^\pi$ and energy $E$ as
\begin{equation}
\hat{\psi}^{J^\pi}_E(r)=\int r'^2 dr' N^{1/2}(r,r')\ \braket{\Phi(r')}{J^\pi\!,E} \ \ \
{\rm with~the~RGM~norm~kernel}\ \ \
N(r,r')=\braket{\Phi(r)}{\Phi(r')} \ .
\end{equation}

Fig.~\ref{fig:amplitudes} shows three $J^\pi$=1/2$^+$ scattering overlap functions
for different energies and the one for the $J^\pi$=3/2$^-$ ground state.
The dashed lines are the Whittaker function and the phase shifted Coulomb wave functions 
for scattering and bound states, respectively. 
When there is no nuclear interaction between the clusters the overlap function has
to go over into the Coulomb wave functions.
The deviation from the Coulomb solution indicates that the nuclear interaction acts up to 10~fm 
in the scattering states up to about 6~fm in the more compact ground state. 
The ground state is an $\ell$=1 state with one node, while the 
$\ell$=0 scattering states are suppressed in the interior due to the Pauli principle
and have two nodes. The $\ell$=0 states with zero and one node are Pauli forbidden
and do not exist.

\begin{figure}[t]
  \includegraphics[width=0.25\columnwidth]{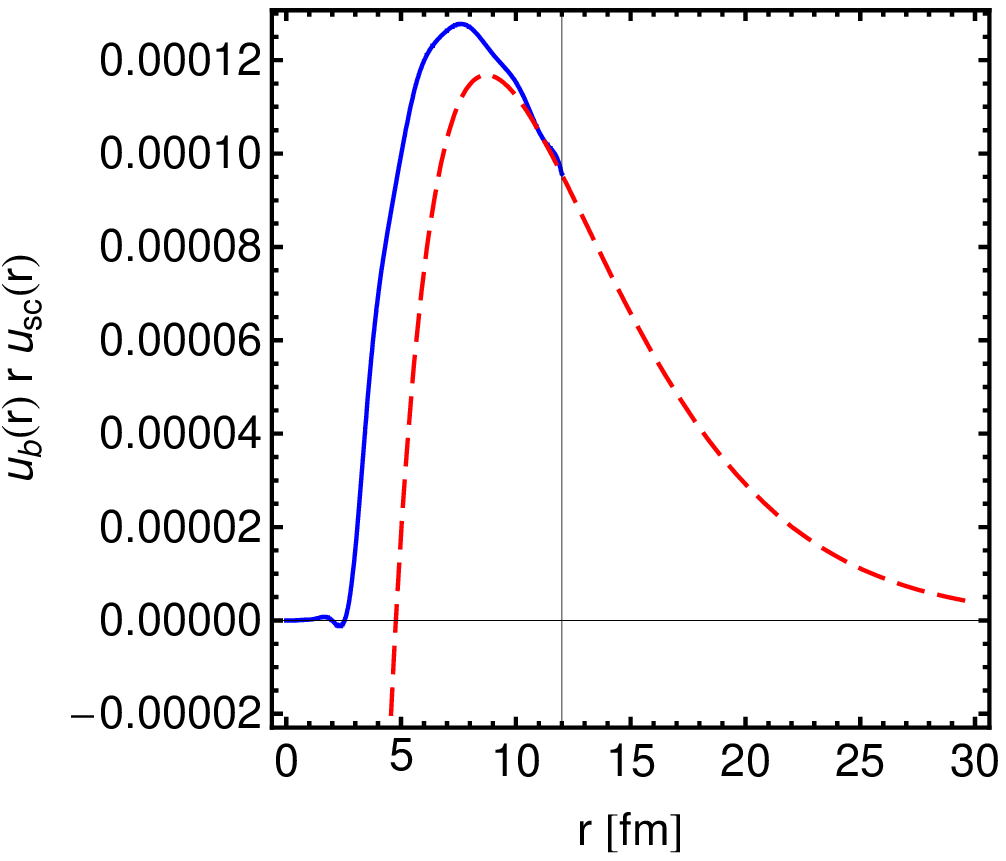}\hfill
  \includegraphics[width=0.245\columnwidth]{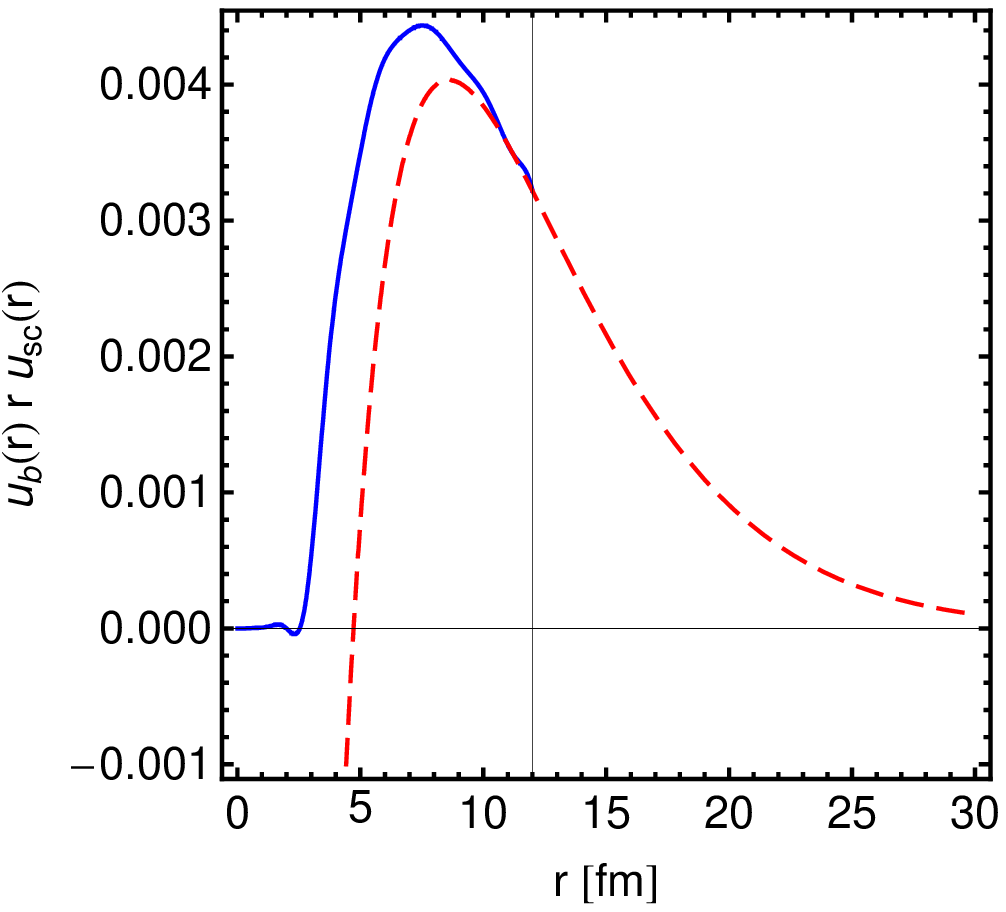}\hfill
  \includegraphics[width=0.237\columnwidth]{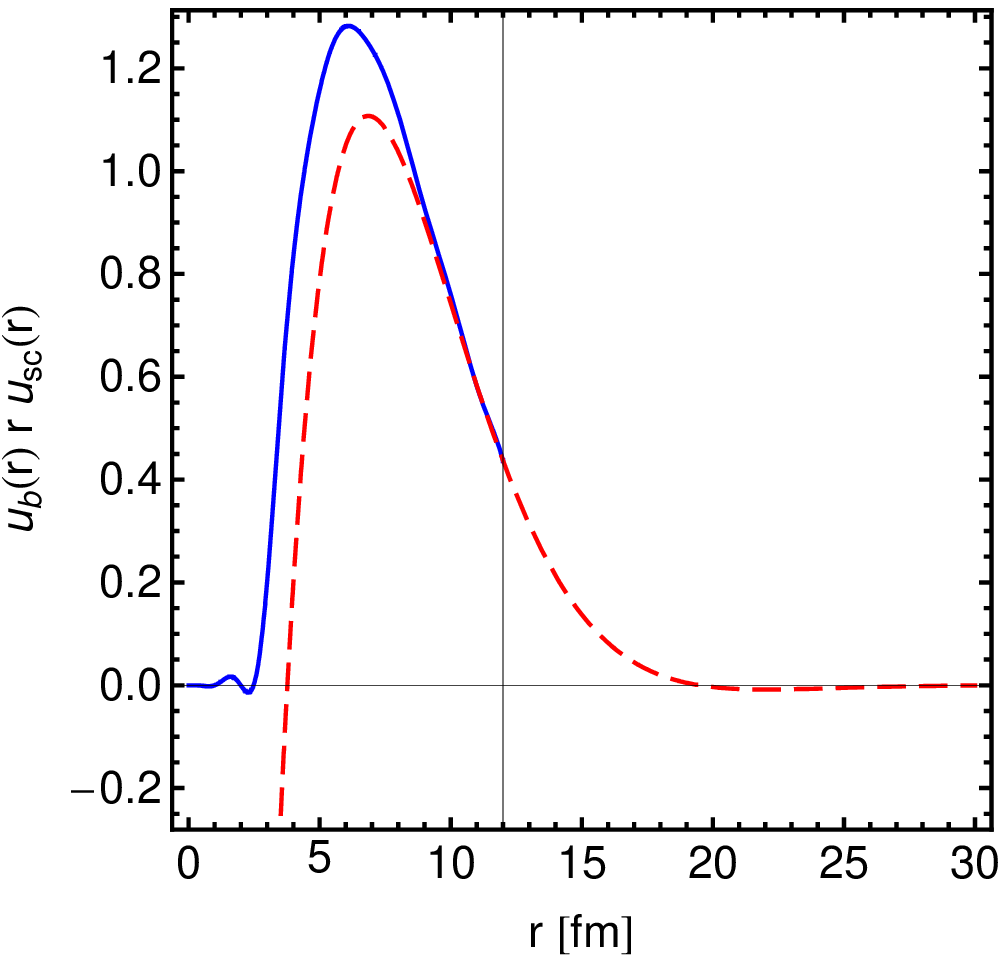}\hfill
  \includegraphics[width=0.237\columnwidth]{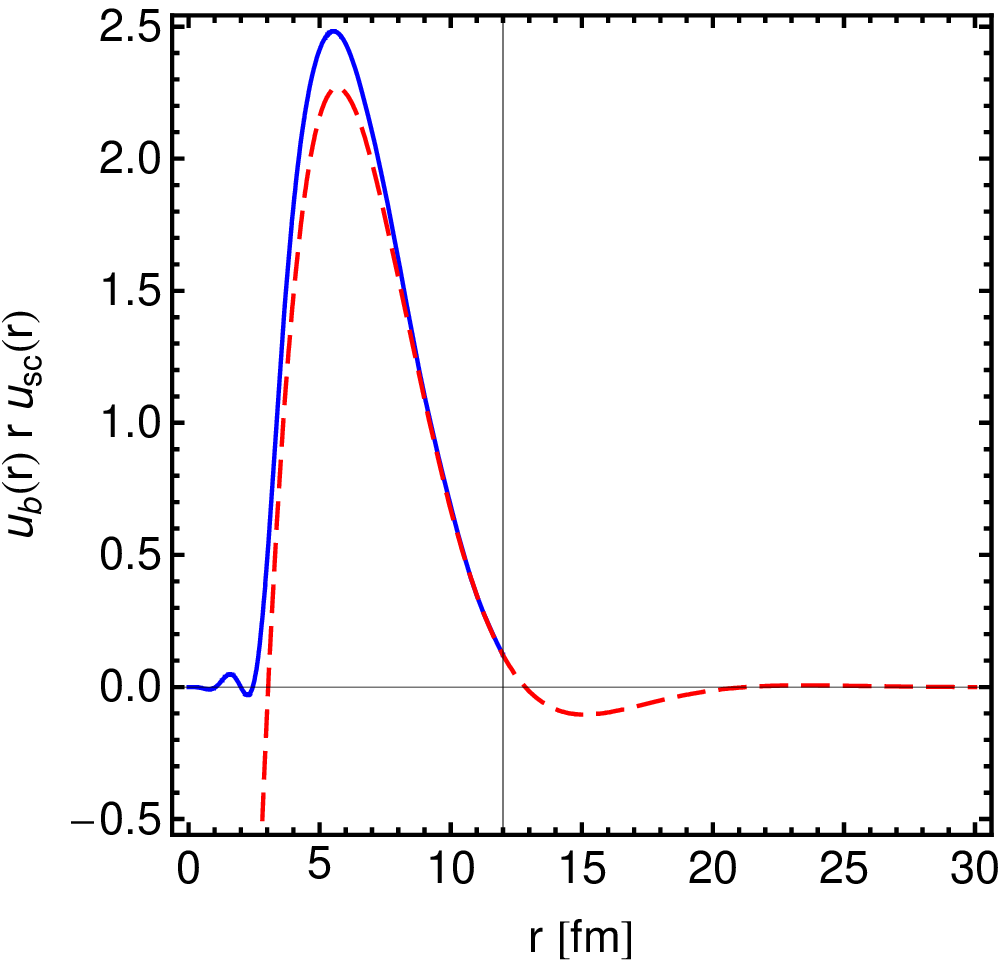}
  \caption{Dipole strength 
  $r\hat{\psi}^{3/2^-}_{gs}\!(r)\cdot r \cdot r\hat{\psi}^{1/2^+}_E\!\!(r)$
  between $3/2^-$ bound state and $1/2^+$ scattering states with energies of  0.05, 0.1, 1.0 
  and 2.0~MeV. Dashed lines for Whittaker and phase shifted Coulomb functions
  matched at 12~fm.}
  \label{fig:dipolemes}
\end{figure}

When looking at the dipole strength displayed in Fig.~\ref{fig:dipolemes}
it can be seen that at low energies a large fraction of the dipole matrix element comes
from the external region where only the Coulomb force acts. 
But it should also be noted that even at energies as low as 50 keV we have a significant 
contribution from the interior region, contrary to common believe that external capture is 
a good approximation at low energies.

\section{Capture Cross Section}
At low energies we can restrict ourselves to electric dipole transitions from the 
$S$- and $D$-wave channels.
\begin{figure}[h]
  \centering
  \includegraphics[width=0.45\textwidth]{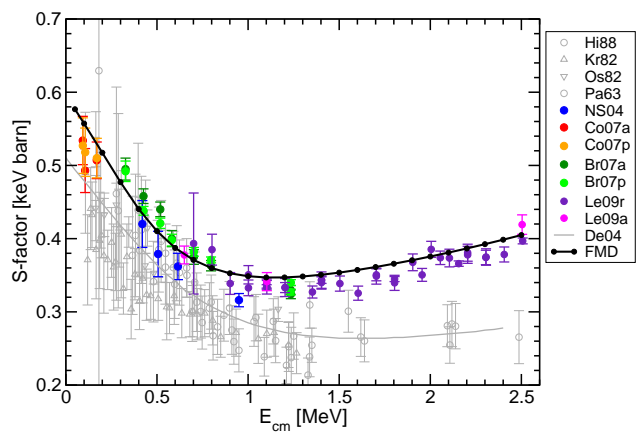}
  \caption{The astrophysical $S$-factor for the \heag\/ reaction. The FMD result is given 
           by the black solid line.  Recent data are shown by dark (colored) symbols.
           Old data are included as gray symbols together with an $R$-matrix fit.}
  \label{fig:sfactor}
\end{figure}
Using the microscopic bound and scattering wave functions and adding up all contributions 
from the $J^\pi=1/2^+,3/2^+,5/2^+$ scattering states to the
$J^\pi= 3/2^-,1/2^-$ bound states
we calculate the radiative capture cross section which is presented in form of the 
astrophysical $S$-factor in Fig.~\ref{fig:sfactor}. 
It agrees very well, both in absolute normalization and energy dependence, 
with the recent experimental results obtained at the 
Weizmann Institute \cite{narasingh04}, the LUNA collaboration \cite{confortola07}, 
at Seattle \cite{brown07} and by the ERNA collaboration \cite{dileva09}. 
One has to keep in mind that in this {\it ab initio} approach there are no parameters
that can be adjusted to the data. Input are the effective realistic nucleon-nucleon
interaction and a suitably chosen many-body Hilbert space.

\begin{figure}[h]
	\centering
  \includegraphics[width=0.45\textwidth]{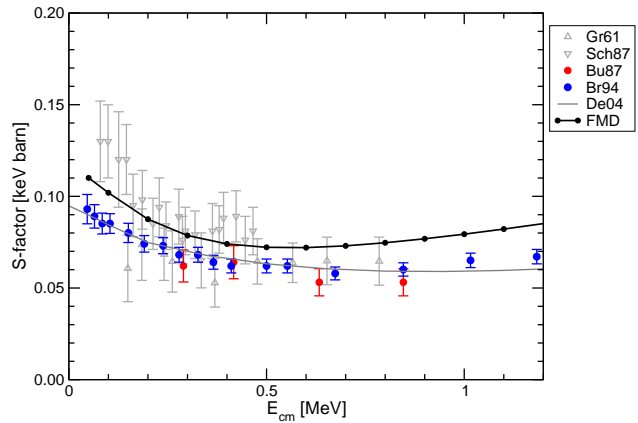}
  \caption{The astrophysical $S$-factor for the \nuc{H}{3}($\alpha$,$\gamma$)\nuc{Li}{7}
     reaction. Recent data are shown by dark (colored) symbols together with an $R$-matrix fit.
     The FMD result is given by the black solid line. Old data are included as 
     gray symbols. }
	\label{fig:s2factor}
\end{figure}

We also did the corresponding calculation for the
\nuc{H}{3}($\alpha$,$\gamma$)\nuc{Li}{7} capture cross sections and found 
that the measured data are about 15\% lower than our calculation, see Fig.~\ref{fig:s2factor}.

Future investigations of our results have to reveal why the microscopic FMD model together
with a realistic nuclear force describes the data while other models fail. 
Compared to previous works there are two main differences, the NN-interaction
and the many-body Hilbert space. As effective interaction and Hilbert space are strongly 
correlated we do not expect to identify a single simple reason, like lacking momentum 
dependence in potential models or too simplified NN-interactions in the microscopic 
cluster models. It is however clear that on has to include polarized clusters 
and shell model like $^7$Be ($^7$Li) states in order to get a successful picture. 
This brings about the very interesting and general question how composite polarizable 
many-body systems perform quantum tunneling.  

$^\dagger$ Contribution to the proceedings of the International School of Nuclear Physics, 
32nd course, Particle and Nuclear Astrophysics, Erice Sicily, September 16 - 24, 2010.\\
To be published in Progress in Nuclear and Particle Physics, Ed. Amand F\"a{\ss}ler.

\end{document}